# Optimal Rate for Irregular LDPC Codes in Binary Erasure Channel

H. Tavakoli
Electrical Engineering Department
K.N. Toosi University of
Technology, Tehran, Iran
tavakoli@ee.kntu.ac.ir

M. Ahmadian Attari
Electrical Engineering Department
K.N. Toosi University of
Technology, Tehran, Iran
m_ahmadian@kntu.ac.ir

M. Reza Peyghami
Department of Mathematics
K.N. Toosi University of
Technology, Tehran, Iran
peyghami@kntu.ac.ir

*Abstract*—In this paper, we introduce a new practical and general method for solving the main problem of designing the capacity approaching, optimal rate, irregular low-density parity-check (LDPC) code ensemble over binary erasure channel (BEC).
Compared to some new researches, which are based on application of asymptotic analysis tools out of optimization process, the proposed method is much simpler, faster, accurate and practical. Because of not using any relaxation or any approximate solution like previous works, the found answer with this method is optimal. We can construct optimal variable node degree distribution for any given binary erasure rate, $\varepsilon$, and any check node degree distribution. The presented method is implemented and works well in practice. The time complexity of this method is of polynomial order. As a result, we obtain some degree distribution which their rates are close to the capacity.

*Keywords-component;LDPC code, Infinite analysis method, Density evolution, LP, SDP*

## I. INTRODUCTION

One of the powerful error-correcting codes, Low-Density Parity-Check (LDPC) codes introduced originally by Gallager in the early sixties [1]. After introduction of Turbo Codes [2], the MacKay's rediscovery cause some excitedness to the LDPC codes [3]. One important property of long irregular LDPC code is achieving and approaching the capacity of the channel [4]. Binary Erasure Channel (BEC), one of the basic channels, has been studied for achieving and approaching the channel capacity [5]. BEC is usually used as a test problem for the capacity approaching problem, because its constraint, which comes from Density Evolution (DE), is simpler than other type of channel. DE, which was first introduced by Richardson and Urbanke in [6] and extended in [4] and [7], is a numerical tool to show how the iterative message passing decoder, the main LDPC decoder, and works with infinite number of iterations. Both asymptotical consideration of infinite number of iteration, and finite length LDPC codes over BEC has been studied in [8-10].

Nowadays, well known methods for constructing these codes are introduced that approach the channel capacity [11], [12] and [7].

However, a method for achieving the capacity in basic channels has still remained as an open problem. Even more, there is no efficient method to generate practical capacity approaching degree distribution for the BEC. Although some experiences for achieving the channel capacity in infinite node degree distribution were done [13-15].

In [16], a family of optimization problems for finding some LP bounds on the degree distributions over memoryless binary-input output-symmetric (MBIOS) channels was introduced. Instead of solving the non-linear constraint, the main problem of capacity approaching presented in Section II of this paper, using other way such as information-theoretic bounds or a method based on error probability and gap to the capacity, for finding LP constraints was replaced. But finding good degree distribution still remains as a hard problem. However, linear programming tool is mostly used to find an optimum solution for the combinatorial optimization problems. One of the most effective usages is LP decoder introduced by Feldman [17].

Optimization models for finding good degree distribution can be classified in three categories. In the first category, algorithm is based on an evolutionary optimization method such as; hill climbing, genetic algorithm and so on [4, Section IV.]. It is known that the Evolutionary Optimization Solvers (EOS) suffer from some disadvantages such as: 1) EOS cannot guarantee the feasible answer. 2) EOS, most of the time, do not converge. 3) Performance of EOS is sensitive to their subroutine and starting point. These disadvantages restrict EOS to be used in wider classes of problems.

In the second category, algorithm is based on an optimization method, such as Differential Evolution, to search a direct way towards the answer [4, Section IV.]. These types of algorithms are based on infinite number of iterations which may have a loop without consideration of convergence or certification for the optimal answer.

In the third category, for solving the capacity approaching problem, using LP method is the main approach. In this manner, a heuristic approach, which is not efficient to find good degree distribution has been used in [8, Section V.]. In this method, for maximizing the rate, minimizing the $\sum \rho_j / j$ is considered. So the necessary condition is $\rho(1 - \varepsilon\lambda(x)) > 1 - x$ on $[0,1]$, where $\rho$ and $\lambda$ are check and variable nodes degree distributions, respectively. In this approach for designing good codes, finding $\rho_j$s is the main problem provided that $\lambda_i$s are given. So the optimization problem is:

Min $\sum \frac{\rho_j}{j}$

Subject to: $\sum \rho_j(1 - \varepsilon\lambda(x_i)) > 1 - x_i$

where $x_i$s, $1 < i < N$, are a set of some fractional values in [0,1]. So an LP problem for finding $\rho_j$s can be defined.

Other methods for designing good degree distribution are presented in [18] and [19]. The basic idea of this way is reformulation of the inequality $\rho(1 - \varepsilon\lambda(x)) > 1 - x$ based on Taylor's series of $\rho^{-1}(1 - x)$ which gives an infinite series. The variable degree sequences gives by this series.

In contrast to these methods, which have been studied, our method is based on the exact constraint with no relaxation. By relaxation the optimization problem the answer would be sub-optimal. So, by this consideration it is certified that the answer of our SDP problem would be optimal.

The organization of the paper is as follows. In Section II, we provide a brief background on problem definition the main problem for optimizing degree distribution. In Section III, we describe SDP reformulation for optimal rate problem. In Section IV, we introduce how we can optimize a code. At last in Section V, we illustrate our contribution with simulation result.

## II. PROBLEM DEFINITION

In this section, we focus on irregular LDPC code ensemble over BEC channel. Let G be a bipartite graph with k message bits that is chosen at random with edge degrees specified by two polynomials:

$$\rho(x) = \sum_{j=2}^{D_c} \rho_j x^{j-1} \qquad \lambda(x) = \sum_{i=2}^{D_v} \lambda_i x^{i-1} \qquad (1)$$

Where $D_c$ and $D_v$ are maximum check and variable node degrees, respectively, and the coefficients of both polynomials are probabilistic, i.e.,

$$\sum_{j=2}^{D_c} \rho_j = 1 \qquad \sum_{i=2}^{D_v} \lambda_i = 1 \qquad \lambda_i \geq 0, \rho_j \geq 0 \qquad (2)$$

Let $\overline{d_c} = 1/\left(\sum_{j=2}^{D_c} \rho_j/j\right)$ and $\overline{d_v} = 1/\left(\sum_{i=2}^{D_v} \lambda_i/i\right)$ denote the average check and average variable nodes, respectively. It is well known that the code rate is defined as [7]:

$$R = 1 - \frac{\overline{d_v}}{\overline{d_c}} \qquad (3)$$

For given erasure probability $\varepsilon > 0$ in BEC, the related channel capacity will be $C = 1 - \varepsilon$. In BEC with given degree distribution on codes, the necessary and sufficient condition for achieving the zero error probability for erased bits, which comes from DE, is

$$\lambda(1 - \rho(1 - x)) \leq \frac{x}{\varepsilon} \quad \forall x \in [0, \varepsilon] \qquad (4)$$

Now, suppose that the check node degree is fixed. Then, in order to maximize the rate of the code, while achieving zero error probability, it is sufficient to solve the following optimization problem:

Max $\sum \frac{\lambda_i}{i}$ (5)

Subject to: $\lambda_i \geq 0$

$\sum \lambda_i = 1$

$\sum \lambda_i (1 - \rho(1 - x))^{i-1} \leq \frac{x}{\varepsilon} \quad \forall x \in (0, \varepsilon]$

This problem is a semi-infinite optimization problem, i.e., it includes infinite number of constraints. It is notable that the rate of the code is not constrained while general formulation mentioned. One way for solving this problem is to discretize the problem by partitioning the continuous interval $(0, \varepsilon]$ for x to discreet set $\{x_0, x_1, ..., x_N\} \subseteq (0, \varepsilon]$ [8, Section V.]. In this case the problem is converted to an optimization problem with finite number of constraints, but the cost that has been paid for this discretization is that a sub-optimal solution is achieved. It is worth mentioning that this optimization problem is linear with respect to the unknown values $(\lambda_1, \lambda_2, ..., \lambda_{D_v})$.

Considering the above argument, one can realize that the main problem in Eq. (5) is the last constraint. Due to this fact, in this paper, we are going to reformulate this constraint as a Linear Matrix Inequality (LMI) and therefore we get a semidefinite reformulation for Eq. (5). By doing this reformulation, we are able to solve it by using polynomial time interior-point methods. It is notable that this reformulation leads to an exact solution for solving the problem instead of suboptimal solutions. Numerical results confirm our claim in comparison with the existence method.

## III. SDP REFORMULATION FOR THE OPTIMAL RATE PROBLEM

### A. Problem Formulation

Our aim in this section is to reformulate the infinitely many constrained optimization problem Eq.(5) as an equivalent semidefinite programming problem in order to apply the well known polynomial time interior-point methods to solve this problem.

Let us briefly describe our reformulation. We first restrict ourselves to discuss about the main constraint of the problem, i.e.,

$$\lambda(1 - \rho(1 - x)) \leq \frac{x}{\varepsilon} \quad \forall x \in [0, \varepsilon] \qquad (6)$$

or about its equivalent reformulation as

$$\lambda(1 - \rho(1 - \varepsilon x)) \leq x \quad \forall x \in [0, 1] \qquad (7)$$

In other words, the feasible region of the problem (5) contains all the vectors $\lambda$ that satisfy in the following relation:

$$P(x) = x - \lambda(1 - \rho(1 - \varepsilon x)) \geq 0, \quad \forall x \in [0, 1] \qquad (8)$$

Since $\lambda(x)$ and $\rho(x)$ are polynomials, so the function $P(x)$ a polynomial function with degree at most $D_c D_v$. Let $q = D_c D_v$ and

$$P(x) = \sum_{j=1}^{q} p_j x^j \qquad (9)$$

where $p_j = p_j(\lambda_1, \lambda_2, \ldots, \lambda_{D_v}, \rho_1, \rho_2, \ldots, \rho_{D_c}, \varepsilon)$. Using Binomial Theorem and some simple calculations, one can verify that for $\rho(x) = x^n$, the coefficients $p_j$ are computed as follows:

$$p_j = \begin{cases} 1 - \sum_i \lambda_i \varphi_{k,i-1} & j = 1 \\ -\sum_i \sum_{l=2}^{j+1} \lambda_l \varphi_{j,l-1} x^k & j \neq 1 \end{cases} \qquad (10)$$

where the coefficients $\varphi_{k,i-1}$ are defined as:

$$\varphi_{k,i-1} = (-1)^{k+i} \varepsilon^k \sum_{\pi_l \geq 0: \sum_{l=1}^{i-1} \pi_l = k} \binom{n}{\pi_1}\binom{n}{\pi_2}\cdots\binom{n}{\pi_{i-1}}$$

In order to express Eq. (8) as an LMI, we first focus on the reformulation of its general form, i.e.,

$$P(x) \geq 0, \qquad \forall x \in \mathbb{R}. \qquad (11)$$

and assume that $q=2k$ is an even number. It has been proved in [20] that the infinitely many constraints (11) is semi-definite representable and its semi-definite representation is:

$$\{\lambda | P(x) \geq 0, \qquad \forall x \in \mathbb{R}\} =$$
$$\{\lambda | \exists B \in S_+^{k+1}; \ p_l = \sum_{i+j=l} B_{ij}, \quad \forall 0 \leq l \leq q = 2k\} \quad (12)$$

Where $S_+^{k+1}$ denotes the set of all symmetric matrices of order $k+1$ that are positive semi-definite. Using the fact that the affine mapping and its image retain the semi-definite representability of sets, one can extend the above mentioned representation to the set $\{\lambda | P(x) \geq 0, \forall x \in [0, \infty)\}$ by the of the affine map $P(x) \to P(x^2)$ from $\mathbb{R}$ to $[0, \infty)$. Similarly, using the affine map $P(x) \to (1+x^2)^q P\left(\frac{x^2}{1+x^2}\right)$ from $\mathbb{R}$ to $[0,1]$, one can easily see that the set

$$\{\lambda | P(x) \geq 0, \qquad \forall x \in [0,1]\}$$

is a semi-definite representable set. In the next section, we provide a semi-definite representation for this set.

## IV. CODE OPTIMIZATION

In this section we provide an explicit semi-definite representation for the set $\{\lambda | P(x) \geq 0, \forall x \in [0,1]\}$. In fact, we would like to replace the infinitely many constraints Eq. (8) by the intersection of affine constraints and some finite LMIs. This leads us to solve the problem (5) taking full responsibility to the all constraints instead of ignoring some of these constraints by discretizing the interval [0, 1] to finite points, as it has been considered in the literature [8]. The following results lead us to the aim of this section.

**Lemma1**: Let $\Pi(x) = (1+x^2)^q P\left(\frac{x^2}{1+x^2}\right) = \sum_{j=0}^{2q} \Pi_j x^j$, where $P(x)$ is defined as Eq. (11). Then we have:

$$\Pi_t = \begin{cases} \sum_{i=1}^{j} \binom{q-i+1}{j-i+1} p_{i-1} & t = 2j \\ 0 & t = 2j+1 \end{cases} \qquad (13)$$

**Proof**: We have:

$$\Pi(x) = \sum_{j=0}^{q} p_j x^{2j}(x^2+1)^{q-j} \qquad (14)$$

Using Newton's expansion, we obtain:

$x^{2j}(x^2+1)^{q-j} = x^{2j} \sum_{r=0}^{q-j} \binom{q-j}{r} x^{2r} = \sum_{r=0}^{q-j} \binom{q-j}{r} x^{2r+2j} = \binom{q-j}{0} x^{2j} + \binom{q-j}{1} x^{2+2j} + \binom{q-j}{2} x^{4+2j} + \cdots + x^{2q}$

Therefore,

$\Pi(x) = \sum_{j=1}^{q} \left\{ p_j \binom{q-j}{0} x^{2j} + p_j \binom{q-j}{1} x^{2+2j} + p_j \binom{q-j}{2} x^{4+2j} + \cdots + p_j x^{2q} \right\}$

This easily shows that

$$\Pi_t = \begin{cases} \sum_{i=1}^{j} \binom{q-i+1}{j-i+1} p_{i-1} & t = 2j \\ 0 & t = 2j+1 \end{cases}$$

which completes the proof of the lemma.

**Theorem1**: Let $\Pi(x)$ be defined as in Lemma 1. Then, the problem (5) is equivalent to the following semi-definite programming problem:

Max $\sum \frac{\lambda_i}{i}$

Subject to: $\sum \lambda_i = 1$
$\Pi_l = \sum_{i+j=l} B_{ij}, \qquad 0 \leq l \leq 2q$
$B \succcurlyeq 0, \ 0 \leq \lambda_i \leq 1$

where $\geq$ is the component-wise order on the vectors and $\succcurlyeq$ denotes the Lowner partial order on symmetric matrices that stands for positive semi-definiteness of the matrices.

**Proof**: According to the discussions of pervious section, the vector $\lambda$ satisfies Eq. (7) if and only if its image by affine mapping $P(x) \to \Pi(x) = (1+x^2)^q P\left(\frac{x^2}{1+x^2}\right)$ from $\mathbb{R}$ to $[0,1]$ satisfies $\Pi(x) \geq 0$, for all $x \in \mathbb{R}$. Using (12), this equality happens if and only if there exists a symmetric positive semi-definite matrix $B = (B_{ij})_{(q+1)\times(q+1)}$ so that it satisfies the following equations:

$$\begin{cases} \Pi_l = \sum_{i+j=l} B_{ij}, & 0 \leq l \leq 2q \\ B \succcurlyeq 0, \end{cases}$$

The proof is completed by replacing these system of linear equations and LMIs in the problem (5).

In order to illustrate these results, we provide a simple structure example to show how these results can be handled in the real problems and computer programming.

**Example1**: Suppose that we are looking for the maximum value of the parameter b so that the polynomial function $f(x) = ax^2 + bx + c$ be nonnegative on the interval [0, 1] under the condition $a = c = 1$. We apply the above mentioned

results and we first obtain the coefficients of the function $\Pi(x)$ using Lemma 1 as follows:

$$\Pi(x) = (1 + x^2)^2 f\left(\frac{x^2}{1 + x^2}\right)$$
$$= (a + b + c)x^4 + (b + 2c)x^2 + c$$
$$= (2 + b)x^4 + (b + 2)x^2 + 1$$

Using Theorem 1, the equivalent semi-definite programming reformulation of the problem is defined as follows:

Max b

Subject to: $y_2 = 1$

$y_3 + y_5 = 0$

$-y_1 + y_4 + y_6 + y_8 = 2$

$y_7 + y_9 = 0$

$-y_1 + y_{10} = 2$

$y_1 = b$

$$\begin{bmatrix} y_2 & y_3 & y_4 \\ y_5 & y_6 & y_7 \\ y_8 & y_9 & y_{10} \end{bmatrix} \succeq 0$$

Using SDP softwares, such as SeDuMi and CVX, lead us to optimal solution $b = 1$, which can be verifies also by using classical solution ways such as interior point method [20].

## V. SIMULATION RESULTS

In this part of paper, we present some numerical results obtained by computer simulations. In this simulation, regular parity check node degree distributions are considered, although an example of using irregular parity check node degree distributions is also presented.

TABLE I. NUMERICAL RESULTS FOR RATE MAXIMIZATION BY REGULAR PARITY CHECK NODE

|  | $\rho(x) = x^3$ | $\rho(x) = x^4$ | $\rho(x) = x^5$ | $\rho(x) = x^6$ | $\rho(x) = x^7$ |
|---|---|---|---|---|---|
| $\lambda_2$ | 0.4735 | 0.4393 | 0.4021 | 0.4385 | 0.4329 |
| $\lambda_3$ | 0.2244 | 0.2097 | 0.2137 | 0.1456 | 0.1583 |
| $\lambda_4$ | 0 | 0.0536 | 0 | 0 | 0 |
| $\lambda_5$ | 0 | 0.2974 | 0 | 0.4159 | 0.4088 |
| $\lambda_6$ | 0 | 0 | 0 | 0 | 0 |
| $\lambda_7$ | 0.3021 | 0 | 0.3902 | 0 | 0 |
| $\varepsilon$ | 0.69 | 0.56 | 0.49 | 0.38 | 0.33 |
| $\varepsilon^{th}$ | 0.69 | 0.56 | 0.49 | 0.38 | 0.33 |
| R | 0.2959 | 0.421 | 0.4922 | 0.593 | 0.6439 |
| C | 0.31 | 0.44 | 0.51 | 0.62 | 0.67 |
| $\delta$ | 0.0478 | 0.0432 | 0.0349 | 0.0435 | 0.039 |

**Example2**: If $\rho(x) = 0.48555x^5 + 0.51445x^6$, the best variable degree distribution with 6 degree is $\lambda(x) = 0.4032x + 0.1512x^2 + 0.4454x^6$. Corresponding rate is Rate = 0.5267 with Capacity = 0.55.

In following, some results for comprising with Table.1 Presented. Our criteria for comparing are based on 5 criteria which presented in [7]. They are:

1-lower maximum degree

2-high rate

3-high threshold

4-lower fraction of degree-two edges

5-ratio $\delta = 1 - R/C$

First of all, in [19] For a Type-A, a code with $\varepsilon = 0.48$ and $\rho(x) = x^5$ is introduced, the variable node degree distribution is:

$$\lambda(x) = 0.4167x + 0.1667x^2 + 0.1000x^3 + 0.0700x^4$$
$$+ 0.0532x^5 + 0.0426x^6 + 0.0353x^7$$
$$+ 0.0300x^8 + 0.0260x^9 + 0.0229x^{10}$$
$$+ 0.0204x^{11} + 0.0165x^{12}$$

For this code we have $R = 0.4998$ and $\varepsilon^{th} = 0.48$. If we use the ratio between rate and capacity we have $R/C = 0.9611$.

In continue, for a Type-MB in [19] a code with $\varepsilon = 0.48$ and $\rho(x) = x^5$ is introduced, the variable node degree distribution is:

$$\lambda(x) = 0.4167x + 0.1667x^2 + 0.1000x^3 + 0.3176x^7$$

For this code we have $R = 0.4926$. If we use the ratio between rate and capacity we have $R/C = 0.9473$.

Secondly, in [21] a method is represented which used some LP solver after each other, one of the best answer with this method is a code with $\varepsilon = 0.5$ and $R = 0.433942$

$$\lambda(x) = 0.205031x + 0.455716x^2 + 0.193248x^{13}$$
$$+ 0.146004x^{14}$$

$$\rho(x) = 0.608291x^5 + 0.391709x^6$$

If we use the ratio between rate and capacity we have $R/C = 0.86$. For an example in related database as a code, we have with $\varepsilon = 0.948$ and $R = 0.05$

$$\lambda(x) = 0.553245x + 0.154015x^2 + 0.0169892x^3$$
$$+ 0.11779x^4 + 0.0673477x^8$$
$$+ 0.0163603x^9 + 0.074253x^{20}$$

$$\rho(x) = 0.1x + 0.9x^2$$

If we use the ratio between rate and capacity we have $R/C = 0.961$.

Thirdly, there is an optimized code in [7.Example3.63] with $\varepsilon = 0.4741$ and $R = 0.5$

$$\lambda(x) = 0.106257x + 0.486659x^2 + 0.010390x^{10}$$
$$+ 0.396694x^{19}$$

$$\rho(x) = 0.5x^7 + 0.5x^8$$

If we use the ratio between rate and capacity we have R/C = 0.9507.

In order to show the efficiency and effectiveness of our proposed, we discretize the interval (0,1] into a discreet set $\{x_0, x_1, \ldots, x_N\}$, for different N's. The results are given in Fig. 1 in which the horizontal axe shows N and the vertical axe is the obtained values of $\lambda_i$'s.

Fig.1 FOUND ANSWER BY USING DESCRETIZING.

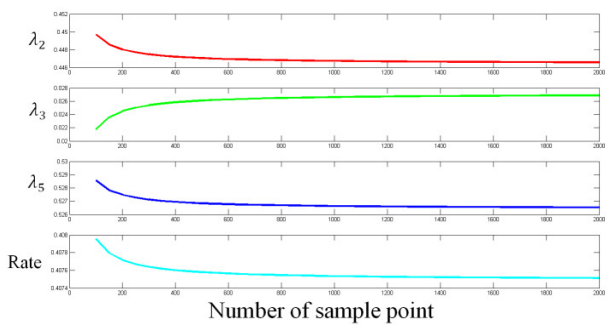

Comparing to the best results obtained until now, the found answers are better in rate maximization based on 5 criteria.

## VI. CONCLUSION

In this paper, first of all, we showed how rate optimization problem as an NLP, can be modeled as a semidefinite problem without any relaxation or simplification. Simulation results in both regular and irregular parity check node degree distribution were presented. These results in most cases are better than the best reported results.

### ACKNOWLEDGEMENT

This work was partially supported by ITRC.